\documentclass[twocolumn,showpacs,prb]{revtex4}
\usepackage{graphicx}
\usepackage{dcolumn}
\usepackage{amsmath}
\usepackage{amssymb}

\makeatletter
\def\btt#1{\texttt{\@backslashchar#1}}%
\DeclareRobustCommand\bblash{\btt{\@backslashchar}}%
\makeatother

\begin{document}

\title[Short Title]{
Spin Current in $\boldsymbol{p}$-wave Superconducting Rings}

\author{Yasuhiro Asano}
\email{asano@eng.hokudai.ac.jp}
\affiliation{%
Department of Applied Physics, Hokkaido University, Sapporo 060-8628, Japan\\
}%

\date{\today}

\begin{abstract}
A formula of spin currents in mesoscopic superconductors is derived from
the mean-field theory of superconductivity.
Spin flow is generated by spatial gradient of $\vec{d}$ which represents a spin state 
of spin-triplet superconductors.
We discuss a possibility of circulating spin currents in isolated 
$p$-wave superconducting rings at the zero magnetic field. 
The direction of spin currents depends on topological numbers which characterize
spatial configurations of $\vec{d}$ on the ring.
\end{abstract}

% PACS,
%\keywords{Suggested keywords}%Use showkeys class option if keyword
                              %display desired
\maketitle

The generation and control of spin polarized currents have been the main topics 
in spintronics~\cite{printz,wolf}. 
Spin devices have been originally proposed on ferromagnets or half-metals because
electric currents in such materials basically carry spins at the same time. 
To realize more efficient spin devices, hybrid structures of ferromagnets and 
semiconductors~\cite{datta} may be more advantageous. 
In this research direction, ferromagnetic semiconductors with high Curie temperatures 
are desired to achieve high spin injection rates into semiconductors. 
Ferromagnetism, however, is not always necessary to generate spin currents.
In a recent theoretical study, the dissipationless spin 
current due to the quantum geometric phase was proposed in nonmagnetic 
semiconductors~\cite{murakami}. In contrast to ferromagnetic metals and semiconductors,
superconductors are not major materials in spintronics. 
This might be because Cooper pairs do not carry 
spins in conventional $s$-wave superconductors. 
Cooper pairs in $p$-wave superconductors, however, have the spin degree of freedom.
Thus spin active transport can be expected in spin-triplet superconductors~\cite{maeno}. 

Spin states of Cooper pairs are described by three components of $\vec{d}$.  
Throughout this paper, vectors in spin space are denoted by $\vec{\cdots}$.
Properties of $\vec{d}$ are similar to those of magnetic moments 
$\vec{m}$ in ferromagnets because both of them characterize spin polarization. 
Here we summarize three important differences between $\vec{d}$ and $\vec{m}$. 
Firstly the time-reversal symmetry (TRS) is broken in ferromagnets, 
whereas it is preserved in unitary superconducting states 
( i.e., $\vec{d} \times \vec{d}^\ast=0$).
It is known that breaking down of the TRS is necessary to generate electric current 
in equilibrium.
Secondly $\vec{m}$ is an observable in ferromagnets, whereas 
 $\vec{d}$ itself is nonobservable. This is because $\vec{d}$ is a part of the 
wave function of Cooper pairs.
The fractional-flux states were discussed~\cite{volovik} in twisted 
crystals~\cite{tanda} 
based on this property.
Finally $\vec{m}$ is parallel to spin polarization, while
$\vec{d}$ points the perpendicular direction to spin polarization of Cooper pairs.

In this paper, we derive a formula of spin currents based on the mean-field theory
of superconductivity. To overview basic properties of spin currents, 
we first apply the obtained formula to Josephson junctions 
of spin-triplet superconductors.
Spin currents are represented by the Andreev reflection coefficients, 
which implies that Cooper pairs carry spins. 
 We mainly discuss the circulating spin current in isolated $p$-wave 
superconducting rings by using the obtained formula. 
The direction of spin current depends on spatial configurations of $\vec{d}$ which
are characterized by topological winding numbers. 
Throughout this paper, we take the unit of $\hbar=k_B=c=1$, where $k_B$ is the 
Boltzmann constant and $c$ denotes the speed of light.

Electronic states in superconductors are described by the Bogoliubov-de Gennes equation,
\begin{align}
&\int\!\! d{\boldsymbol{r}'} 
\!\!\left[\begin{array}{cc}
h(\boldsymbol{r},\boldsymbol{r}')\hat{\sigma}_0 &
\hat{\Delta}(\boldsymbol{r},\boldsymbol{r}') \\
- \hat{\Delta}^\ast(\boldsymbol{r},\boldsymbol{r}') &
-h^\ast(\boldsymbol{r},\boldsymbol{r}')\hat{\sigma}_0
\end{array}\right]\!\!
\left[\begin{array}{c}
\hat{u}(\boldsymbol{r}')\\
\hat{v}(\boldsymbol{r}')
\end{array}\right] 
%\nonumber \\
%&
= E \!\!\left[\begin{array}{c}
\hat{u}(\boldsymbol{r})\\
\hat{v}(\boldsymbol{r})
\end{array}\right], \label{bdg}\\
&h(\boldsymbol{r},\boldsymbol{r}')=\delta(\boldsymbol{r}-\boldsymbol{r}')\left\{
-\frac{\boldsymbol{D}_{\boldsymbol{r}'}}{2m} +V(\boldsymbol{r}')-\mu\right\},\label{h1}
\end{align}
where $\boldsymbol{D}_{\boldsymbol{r}}=\boldsymbol{\nabla}_{\boldsymbol{r}}-ie
\boldsymbol{A}_{\boldsymbol{r}}$, $\hat{\cdots}$ indicates $2\times 2$ matrix 
describing spin space, 
$\hat{\sigma}_0$ is the unit matrix
and $\mu$ is the Fermi energy. 
In uniform superconductors, pair potentials are given by
\begin{align}
\hat{\Delta} \left( \boldsymbol{r},\boldsymbol{r}'\right)
=& \frac{1}{V_{vol}} \sum_{\boldsymbol{k}} \hat{\Delta}_{\boldsymbol{k}}
e^{i\boldsymbol{k} \cdot (\boldsymbol{r}-\boldsymbol{r}')}, \\
\hat{\Delta}_{\boldsymbol{k}}=&
 i \; \vec{d} (\boldsymbol{k}) \cdot \hat{\vec{\sigma}} \hat{\sigma}_2
e^{i\varphi},
\end{align}
where $\hat{\sigma}_i$ with $i=1, 2$ and 3 are the Pauli matrices, 
and $\varphi$ is a macroscopic phase.
Using a conservation low of spin density,
we derive a formula of the spin current,
\begin{align}
&\vec{\boldsymbol{J}}_s = \frac{1}{4im}\lim_{\boldsymbol{r}'\to \boldsymbol{r}}
T\sum_{\omega_n}\textrm{Tr} \nonumber\\
\times  & \left[\begin{array}{cc} \boldsymbol{D}_{\boldsymbol{r}} - \boldsymbol{D}^\ast_{\boldsymbol{r}'} &0 \\ 0 &
\boldsymbol{D}^\ast_{\boldsymbol{r}} - \boldsymbol{D}_{\boldsymbol{r}'}\end{array}\right]
\check{\mathfrak{G}}_{\omega_n}(\boldsymbol{r},\boldsymbol{r}')
\frac{1}{2}\left[\begin{array}{cc} \hat{\vec{\sigma}} & \hat{0} \\
\hat{0} & \hat{\vec{\sigma}}^\ast \end{array}\right],\label{sf1}
\end{align}
where $\check{\mathfrak{G}}_{\omega_n}(\boldsymbol{r},\boldsymbol{r}')$ is the Matsubara-Green 
function of Eq.~(\ref{bdg}) in the $4\times 4$ matrix form, 
and Tr is carried out over the Nambu $\times$ spin space.
 When $ \hat{\vec{\sigma}}/2$ 
is replaced by $-e \hat{\sigma}_0$ in Eq.~(\ref{sf1}), 
we obtain the formula of the Josephson electric current 
($\boldsymbol{J}_e$)~\cite{furusaki,asano}.
Eq.~(\ref{sf1}) is a general expression of spin currents which can be applied to any
superconducting systems.
\begin{figure}[htbp]
\begin{center}
\includegraphics[width=8.0cm]{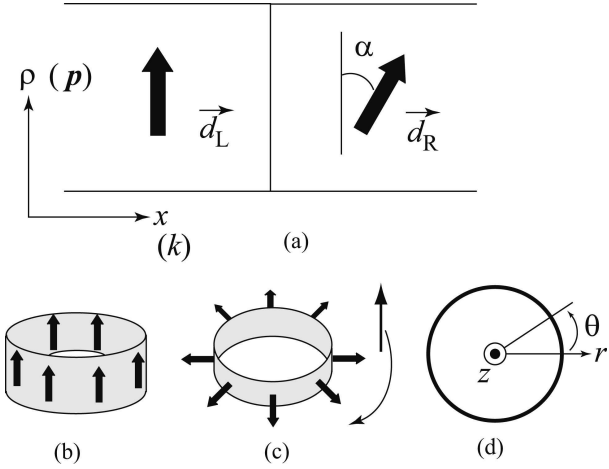}
\end{center}
\vspace{-0.8cm}
\caption{
The Josephson junction under consideration is in (a). 
In (b)-(d), schematic pictures of $p$-wave rings are illustrated.
}
\label{fig1}
\end{figure}
From Eq.~(\ref{sf1}), it is possible to derive a spin current formula
in superconductor / superconductor junctions. 
When triplet superconductors are in unitary states,
spin currents are calculated from the Green function of 
junctions~\cite{asano},
\begin{align}
\vec{\boldsymbol{J}}_s=- &\sum_{\boldsymbol{p}} \frac{T}{4} 
\sum_{\omega_n} \textrm{Tr} \left[
\frac{1}{\Omega_{L,+}}\left\{
\hat{\Delta}_{L,+} \hat{a}_1 \frac{\hat{\vec{\sigma}}}{2}
+  \hat{a}_1 \hat{\Delta}_{L,+} \frac{\hat{\vec{\sigma}}^\ast}{2}\right\}
 \right.\nonumber \\
&\left.
-  \frac{1}{\Omega_{L,-}}\left\{
\hat{a}_2 \hat{\Delta}_{L,-}^\dagger \frac{\hat{\vec{\sigma}}}{2}
+ \hat{\Delta}_{L,-}^\dagger \hat{a}_2 \frac{\hat{\vec{\sigma}}^\ast}{2}
\right\}\right],\label{sf2}
\end{align}
where $\Omega_{L,\pm}=\sqrt{\omega_n^2+|\vec{d}_{L,\pm}|^2}$, 
$\hat{\Delta}_{L,\pm}= i \; \vec{d}_{L,\pm} \cdot \hat{\vec{\sigma}} \hat{\sigma}_2$,
$\vec{d}_{L,\pm} = \vec{d}_L(\pm k, \boldsymbol{p})$ are the vectors in the 
left superconductor. 
In the momentum space,
 $k$ and $\boldsymbol{p}$
are wave numbers on the Fermi surface in the direction of currents and 
in the directions transverse to currents, respectively.
The electric Josephson current is also given by $\hat{\vec{\sigma}}/2 \to -e \hat{\sigma}_0$
in Eq.~(\ref{sf2})~\cite{asano}. 
Most important feature of Eq.~(\ref{sf2}) is that spin currents are represented by 
the two Andreev reflection coefficients
of a quasiparticle incident from a superconductor on the left
hand side, (i.e., $\hat{a}_1$ and $\hat{a}_2$).
This fact implies that Cooper pairs carry spins.

To overview basic properties of spin currents, 
let us briefly discuss spin currents in a Josephson 
junction in the clean limit as shown in Fig.~\ref{fig1}(a).
We assume that $\vec{d}_{L(R)}$ are real vectors satisfying 
$\vec{d}_{L(R),+}=\vec{d}_{L(R),-}=\vec{d}_{L(R)}$ for simplicity. 
We also assume that amplitudes of $\vec{d}$ in two superconductors 
are identical to each other, (i.e., $|\vec{d}_{L}|= |\vec{d}_{R}|=|\vec{d}|$). 
The directions of $\vec{d}_R$ is oriented by $\alpha$ 
from $\vec{d}_L$. 
In the absence of a potential barrier between the two superconductors, 
we first calculate the Andreev reflection coefficients by solving
Bogoliubov-de Gennes equation.
%\begin{align}
%\hat{a}_1=& \frac{i\hat{\Delta}_L^\dagger}{2\omega_n g^\ast}\!\!
%\left[ u^2 \chi^\ast \frac{\hat{\Delta}_L\hat{\Delta}_R^\dagger}{|\vec{d}|^2}-v^2\chi\right]\!\!\!
%\left[ \chi^\ast \frac{\hat{\Delta}_R\hat{\Delta}_L^\dagger}{|\vec{d}|^2}- \chi\right],\\
%\hat{a}_2=& 
%\left[ u^2 \chi - \frac{\hat{\Delta}_L\hat{\Delta}_R^\dagger}{|\vec{d}|^2}v^2\chi\ast\right]\!\!\!
%\left[ \chi - \frac{\hat{\Delta}_R\hat{\Delta}_L^\dagger}{|\vec{d}|^2} \chi^\ast\right]
%\!\!\frac{i\hat{\Delta}_L}{2\omega_n g},\\
%g=&\left(1+\frac{|\vec{d}|^2}{2(\omega_n)^2}\right)\cos\varphi + 
%\frac{|\vec{d}|^2}{2(\omega_n)^2}\cos\alpha \nonumber\\
%& +i \left(\frac{\Omega}{\omega_n}\right) \sin\varphi,
%\end{align}
%where $\chi=e^{i\varphi/2}$ with $\varphi=\varphi_L-\varphi_R$ and $u^2$ ($v^2$) 
%$ = (\omega_n +(-) \Omega)/(2\omega_n)$.  
Electric and spin currents of junctions in Fig.~\ref{fig1}(a) result in, 
\begin{align}
J_e =& e \sum_{\boldsymbol{p}} \frac{|\vec{d}|}{4}
\left[  F_+ + F_-   \right],\label{e1}\\
\vec{J}_s =& \frac{1}{2} \vec{n} \sum_{\boldsymbol{p}} \frac{|\vec{d}|}{4} 
\left[  F_+ - F_- \right],\label{s1} \\
F_\pm =& \frac{\sin\left( \eta_\pm \right)
\cos\left( \eta_\pm \right)}{\left|\cos\left( \eta_\pm \right)\right|}
\tanh\left(\frac{|\vec{d}||\cos(\eta_\pm)|}{2T}\right),
\end{align}
where $\eta_\pm = (\varphi \pm \alpha)/2$,
$\vec{n}=\vec{d}_R\times\vec{d}_L/|\vec{d}_R\times\vec{d}_L|$.
Current-phase relations of charge are always the odd function of $\varphi$
irrespective of $\alpha$.
At $\alpha=0$, the current-phase relation becomes $J_e \propto \sin(\varphi/2)$ at $T=0$
for $-\pi < \varphi < \pi$ as it is in usual short junctions in the clean limit~\cite{ko2}. 
%At $\alpha =\pi/2$, a period of current-phase relations becomes $\pi$.
%In this case, current components proportional to $\sin(m \varphi)$ vanish 
%when $m$ are odd integers. 
From Eq.~(\ref{s1}), it is confirmed that 
the current-phase relation of spin are always the even function of $\varphi$. 
Spin currents are allowed even at $\varphi=0$ (in the presence of TRS). 
Actually at $\varphi=0$, we find 
\begin{align}
\vec{J}_s \propto
\left\{
\begin{array}{ll}
\sin(\alpha/2) &: T=0 \\
\sin(\alpha) &: T\lesssim T_c
\end{array}\right. 
\label{scjj}
\end{align}
for $-\pi < \alpha < \pi$,
where $T_c$ is the critical temperature. It is also confirmed that 
spin currents vanish at $\alpha= \pm \pi$ and 0.
Spin flows are possible when $\vec{d}$ in the two superconductors 
are not aligned in parallel or antiparallel.
The calculated results in Eq.~(\ref{s1}) seem to be consistent with 
those obtained by the quasiclassical Green function theory~\cite{rashedi}.
The spatial part of pair potentials affects amplitudes of spin currents 
through the integration of $|d|$ on the Fermi surface as shown in Eq.(\ref{s1}). 
The difference of pairing symmetries such as $p$- and $f$-wave symmetries is not so important 
in weak links. In the presence of a potential barrier between the two superconductors,
however, properties of spin currents may depend on the spatial part of pair potentials. 
Amplitudes of pair potentials at the interface
may deviate from their bulk values in real materials. Thus it needs to 
determine amplitudes of pair potentials in a self-consistent way in order to
discuss electric and spin currents in realistic junctions. 
%The suppression of pair potentials near the interface may lead to
%suppression of spin currents. 
Such self-consistency of pair potentials is 
not considered here because purpose of this part
is to explain basic properties of spin current by using the analytic
representation in Eq.~(\ref{scjj}).

%The charge and spin currents in Eqs.~(\ref{e1}) and (\ref{s1}) arise no energy dissipation
%because they flow in equilibrium.
%
%\begin{figure}[htbp]
%\begin{center}
%\includegraphics[width=8.0cm]{fig2.eps}
%\end{center}
%\vspace{-0.8cm}
%\caption{
%The current-phase relation in $p$-wave Josephson junctions. 
%The charge current and the spin current are plotted in (a) and (b), respectively. 
%The vertical axes are normalized by appropriate values.
%}
%\label{fig2}
%\end{figure}
%

The results in Eq.~(\ref{scjj}) indicate that spin currents are caused by the 
 spatial gradient of $\vec{d}$.  
Generally speaking, directions of $\vec{d}$ are fixed to underlying lattices of 
superconductors because they are determined by spin anisotropy
due to the spin-orbit coupling. 
Thus the nonhomogeneity of $\vec{d}$ are not expected in usual bulk samples and 
thin films.
The situation, however, would be changed in topological crystals~\cite{tanda} 
as shown in Fig.~\ref{fig1}(b)-(c). 
In (b), no spin current is expected because $\vec{d}$ is homogeneous on the ring.
In (c), however, it is easily found that $\vec{d}$ is rotated by 
$2\pi$ while it circles the ring.
The pair potential in Fig.~\ref{fig1}(c) can be described by
\begin{align}
\hat{\Delta}_{{\theta}}(k,\boldsymbol{p}) = i \vec{e}_r({\theta})
\cdot\hat{\vec{\sigma}}\hat{\sigma}_2 
d(k,\boldsymbol{p}),\label{d-ring}
\end{align}
where $d(k,\boldsymbol{p})$ is the spatial part of the pair potential, 
$\vec{e}_r$ is the unit vector in the $\vec{r}$ direction.
As shown in Fig.~\ref{fig1}(d), wave numbers in $\theta$ and $(r,z)$ directions 
are denoted by
$k$ and $\boldsymbol{p}$, respectively.
We first calculate the Green function of the two-dimensional ring as shown in 
Fig.~\ref{fig1}(c) with Eq.~(\ref{d-ring}).
When the radius of ring is fixed at $r$,
the Green function for $z=z'$ and $\theta \geq \theta'$ are given by
\begin{widetext}
\begin{align}
\check{\mathfrak{G}}_{\omega_n}(\theta,\theta')
=\sum_{\boldsymbol{p}} &\frac{-mr}{2k_F }
\left[
\frac{e^{ik_F(\theta-\theta')}}{\Omega_+^-}
\left( \begin{array}{cc}
(i\omega_n - \gamma +i \Omega_+^-) e^{-i(\theta-\theta')\hat{\sigma}_3/2} \hat{t}_+&
d_+e^{-i(\theta+\theta')\hat{\sigma}_3/2} \hat{t}_+\\
d_+^\ast e^{i(\theta+\theta')\hat{\sigma}_3/2} \hat{t}_+&
(i\omega_n - \gamma -i \Omega_+^-) e^{i(\theta-\theta')\hat{\sigma}_3/2}\hat{t}_+
\end{array}\right) \right. \nonumber \\
&+
\frac{e^{-ik_F(\theta-\theta')}}{\Omega_-^+}
\left( \begin{array}{cc}
(i\omega_n + \gamma -i \Omega_-^+) e^{-i(\theta-\theta')\hat{\sigma}_3/2} \hat{t}_+&
d_-e^{-i(\theta+\theta')\hat{\sigma}_3/2} \hat{t}_+\\
d_-^\ast e^{i(\theta+\theta')\hat{\sigma}_3/2} \hat{t}_+&
(i\omega_n + \gamma +i \Omega_-^+) e^{i(\theta-\theta')\hat{\sigma}_3/2}\hat{t}_+
\end{array}\right)  \nonumber \\
&+
\frac{e^{ik_F(\theta-\theta')}}{\Omega_+^+}
\left( \begin{array}{cc}
(i\omega_n + \gamma +i \Omega_+^+) e^{-i(\theta-\theta')\hat{\sigma}_3/2} \hat{t}_-&
-d_+e^{-i(\theta+\theta')\hat{\sigma}_3/2} \hat{t}_-\\
-d_+^\ast e^{i(\theta+\theta')\hat{\sigma}_3/2} \hat{t}_-&
(i\omega_n + \gamma -i \Omega_+^+) e^{i(\theta-\theta')\hat{\sigma}_3/2}\hat{t}_-
\end{array}\right)  \nonumber \\
&+ \left.
\frac{e^{-ik_F(\theta-\theta')}}{\Omega_-^-}
\left( \begin{array}{cc}
(i\omega_n - \gamma -i \Omega_-^-) e^{-i(\theta-\theta')\hat{\sigma}_3/2} \hat{t}_-&
-d_-e^{-i(\theta+\theta')\hat{\sigma}_3/2} \hat{t}_-\\
-d_-^\ast e^{i(\theta+\theta')\hat{\sigma}_3/2} \hat{t}_-&
(i\omega_n - \gamma +i \Omega_-^-) e^{i(\theta-\theta')\hat{\sigma}_3/2}\hat{t}_-
\end{array}\right)\right],  \label{green}
\end{align}
\end{widetext}
where $\gamma = k_F/2mr^2$, $\hat{t}_\pm=(\hat{\sigma}_0 \pm \hat{\sigma}_3)/2$, $\Omega_{\nu}^\pm=\sqrt{|d_{\nu}|^2 -(i\omega_n \pm \gamma)^2}$
with $\nu= +$ or $-$, 
$d_\pm= d\left( \pm k,\boldsymbol{p}\right)$, 
and $k_F$ is the dimensionless Fermi wave number. 
On the way to Eq.~(\ref{green}), we assume a relation
$\gamma < |d_{+,-}| \ll k_F\gamma$. 
By substituting Eq.~(\ref{green}) into Eq.~(\ref{sf1}), 
spin currents of the ring in Fig.~\ref{fig1} (c) become
at the zero temperature
\begin{align}
\vec{\boldsymbol{J}}_s = (-\boldsymbol{n}_{\theta} )\;
\left( \frac{\vec{e}_z}{2} \right) \; \frac{\pi}{2} \;\gamma
N_c,\label{jsring}
\end{align}
where $\vec{e}_z$ is the unit spin vector in the $z$ direction, 
 $\boldsymbol{n}_{\theta}$ is the unit vector in real space in the $\theta$ direction,
and $N_c$ is the number of propagating channels on the Fermi surface.
We have also assumed that $d_+=d_-\equiv d$.
Under the condition $\gamma < |d|$, the final expression
of spin current in Eq.~(\ref{jsring}) depends only on shapes of rings 
such as diameters and heights. 
Spins polarized in the $+z$ direction flow in clockwise.
In other words, spins polarized in the $-z$ direction flow in counterclockwise.
In the presence of rotation in $\vec{d}$, 
circulating spin currents flow in equilibrium in $p$-wave superconducting rings
because of the non-commutativity of spin algebra. 
In small ferromagnetic rings, it has been pointed out that the non-commutativity 
of spin algebra is a source of the circulating electric current~\cite{tatara}. 

\begin{figure}[htbp]
\begin{center}
\includegraphics[width=8.0cm]{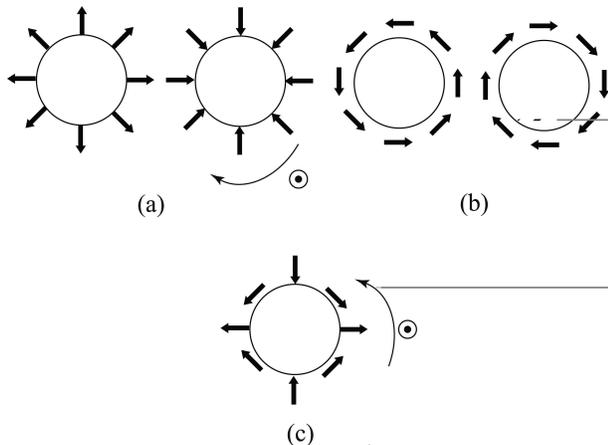}
\end{center}
\vspace{-0.8cm}
\caption{
Various configurations of $\vec{d}$ on topological ring crystals.
The direction of spin currents is the same in four rings in (a) and (b).
In (c), spins flow opposite direction to those in (a) and (b).
}
\label{fig3}
\end{figure}

Directions of spin currents depend on configurations of $\vec{d}$.
In Fig.~\ref{fig3}, we show top views of several $p$-wave rings with 
various configurations of $\vec{d}$.
At first sight, spin currents in the four rings in (a) and (b) seem to be different 
 from one another. The microscopic calculation, however, show that spin currents 
in the four configurations are the same with one another. 
The $\vec{d}$ in the left ring in (a) point opposite directions to those in the right 
ring in (a). The flip of $\vec{d}$ can be described by the $\pi$-phase 
shift in macroscopic phase (i.e., $\vec{d} \to -\vec{d}=\vec{d}e^{i\pi}$).  
It is evident that the uniform phase shift does not affect physical values.
The same argument is valid between the two rings in Fig.~\ref{fig3}(b).
Configurations of $\vec{d}$ in (a) and those in (b) can be related 
to each other by the uniform rotation of $\vec{d}$ by $\pi/2$.
These spatial configurations of $\vec{d}$ on rings are characterized by 
 a topological winding number ($N$).
In the four figures in (a) and (b), $\vec{d}$ is rotated by $2\pi$ while 
$\theta$ increasing from 0 to $2\pi$, (i.e., $N=1$).
Spin currents polarized in $+z$ direction circulate clockwise for $N=1$.
To change the direction of spin currents, we have to consider $\vec{d}$ as shown in 
Fig.~\ref{fig3}(c), where the winding number of $\vec{d}$ becomes $N=-1$.
It is easy to confirm that $\vec{d}$ suffers the rotation by $-2\pi$ while 
$\theta$ increasing from 0 to $2\pi$.
All rings in (a) and (b) show uniform configurations of $\vec{d}$
when we cut a ring at a certain place and deform the ring into the strip. 
In (c), however, the nonhomogeneity in $\vec{d}$ persists even when
the ring is deformed to the strip. 
A possibility of the $\vec{d}$ in (c) is questionable in real 
superconductors because the nonhomogeneity in $\vec{d}$ costs energy.
%The configuration of $\vec{d}$ in (c) is basically 
%impossible as far as the uniform superconductivity is considered on crystal strips.
Therefore spin-up and -down symmetry of Cooper pairs is 
violated in their directions of flow.
Spins polarized $+z$ ($-z$) direction always flow in clockwise (conterclockwise)
on ring crystals of superconductors. 
This property is a new feature of spin-triplet superconductors. 
%Curved strips of triplet superconductors can be a spin selector. 

In summary, we have derived a formula of spin currents in spin-triplet superconducting
systems. Flow of spin in equilibrium is described by the Andreev reflection and is
possible when $\vec{d}$ has spatial gradient.
In $p$-wave superconducting rings, directions of spin currents are determined
by winding numbers which characterize spatial configurations of $\vec{d}$.
The spin-up and -down symmetry in Cooper pair is violated with respect to their 
directions of flow; this is a new feature of spin-triplet superconductors. 
In this paper, we focus on spin-triplet unitary states in the clean limit.
Effects of nonunitary states and those of midgap Andreev resonant 
states~\cite{tanaka} on the spin transport are important open questions.

The author thanks Y.~Maeno, Y.~Tanaka, G.~Tatara and S.~Tanda for useful discussion. 
 This work has been supported by Grant-in-Aid for the 21st Century
COE "Topological Science and Technology" Scientific Research from the Ministry of
Education, Culture, Sport, Science and Technology of Japan.

{}

\end{document}